\documentclass[aps,pra,preprint,superscriptaddress,longbibliography]{revtex4-2}

\usepackage{graphicx}
\usepackage{amsmath}
\usepackage{color}
\usepackage[pdftex]{hyperref} 

\begin{document}
	
	\title{Coherent states of the Laguerre-Gauss modes}

        \author{M.~P. Morales Rodr\'iguez}
        \email[e-mail: ]{mariapilar.morales22@estudiantes.uva.es}
        \affiliation{Departamento de F\'isica Te\'orica, At\'omica y \'Optica, Universidad de Valladolid, 47011 Valladolid, Spain}

        \author{O. S. Magaña-Loaiza}
        \email[e-mail: ]{maganaloaiza@lsu.edu}
        \affiliation{Quantum Photonics Laboratory, Department of Physics \& Astronomy, Louisiana State University,  Baton Rouge, LA 70803-4001}

        \author{B. Perez-Garcia}
        \email[e-mail: ]{b.pegar@tec.mx}
        \affiliation{Photonics and Mathematical Optics Group, Tecnologico de Monterrey, Monterrey 64849, Mexico}

        \author{L.~M. Nieto Calzada}
        \email[e-mail: ]{luismiguel.nieto.calzada@uva.es}
        \affiliation{Departamento de F\'isica Te\'orica, At\'omica y \'Optica, Universidad de Valladolid, 47011 Valladolid, Spain}	
        
        \author{F. Marroqu\'in}
        \email[e-mail: ]{fco.marroquin.gtz@upp.edu.mx}
        \affiliation{Universidad Polit\'ecnica de Pachuca. Carr. Pachuca-Cd. Sahag\'un Km.20, Ex-Hda. Santa B\'arbara. Zempoala, 43830 Hidalgo, Mexico }
        
        \author{B.~M. Rodr\'iguez-Lara}
        \email[e-mail: ]{bmlara@upp.edu.mx}
        \affiliation{Universidad Polit\'ecnica de Pachuca. Carr. Pachuca-Cd. Sahag\'un Km.20, Ex-Hda. Santa B\'arbara. Zempoala, 43830 Hidalgo, Mexico }
	
	\date{\today}

	\date{\today}
	
    \begin{abstract}
        Large quantum photonic systems hold promise for surpassing classical computational limits, yet their state preparation remains a challenge. 
        We propose an alternative approach to study multiparticle dynamics by mapping the excitation mode of these systems to physical properties of the Laguerre-Gauss modes.
        We construct coherent states establishing a direct link between excitation number dynamics and the evolution of the Laguerre-Gauss mode. 
        This highlights the photon transverse spatial degree of freedom as a versatile platform for testing fundamental aspects of quantum multiparticle systems.
    \end{abstract}

    \maketitle
    \newpage

\section{Introduction}

The second quantum revolution seeks to apply the principles of quantum physics to quantum technologies, focusing on unlocking the potential of large multiparticle systems for complex operations through interference and scattering processes \cite{Dowling2003,Schleich2016,Acin2018,Erhard2020}. 
The engineering of robust, controllable quantum systems composed of multiple interacting particles is crucial, and ongoing efforts involve various physical systems like trapped atoms  and ions \cite{Cirac2000,Bloch2008,Duan2010}, superconducting circuits \cite{Kjaergaard2020,Blais2021}, and photonic quantum devices \cite{ Schleich2016,Ott2010,Marty2016,Stobinska2018,Gebhart2023,Browne2017, Pan2012,Raymer2020,Sheremet2023,Cherchi2023}. 
However, the experimental preparation of controllable multiparticle systems faces significant challenges \cite{Kalb2017,Shang2018,Giordani2019,You2021}, necessitating exploration of alternative platforms relying on spatial light modes \cite{Kawase2008, Mafu2013, Malik2015, Tang2016, Krenn2017}. 
This possibility has stimulated classical analogies of entanglement, where the number of particles is replaced by the degrees of freedom of spatial light modes. 
This approach has enabled the exploration of complex multiparticle phenomena through simple manipulation of polarization and spatial properties of classical light beams \cite{Hashemi2015, Ndagano2017, Paneru2020, Qian2015}.

Here, we map the excitation mode of the electromagnetic field to the transverse properties of the spatial Laguerre-Gauss modes. 
To construct coherent states, we leverage symmetries linked to the conservation of quantum numbers and present experimental evidence thereof in multiparticle systems.
We provide a robust platform for exploring the fundamental properties of quantum multiparticle systems harnessing the transverse spatial degree of light.

\section{Results}\label{sec2}

We study the Laguerre-Gauss modes using an operator-based approach, characterized by radial, $p$, and azimuthal, $\ell$, excitation numbers \cite{Allen1992,Calvo2006,Karimi2014,Plick2015},
\begin{align}
    \left\vert  p, \ell \right\rangle =&~ \dfrac{1 }{\sqrt{(p)! \, (p+|\ell|)!}} \left(\hat{a}_{+}^{\dagger} \hat{a}_{-}^{\dagger} \right)^{p} 
    \begin{cases}
		\left(\hat{a}_{+}^{\dagger} \right)^{|\ell|}\left\vert 0 \right\rangle, & \ell>0, \\
		\left(\hat{a}_{-}^{\dagger} \right)^{|\ell|}\left\vert 0 \right\rangle, & \ell<0 ,
	\end{cases}
\end{align}
where the creation operators, $\hat{a}_{\pm}^{\dagger} = ( \hat{a}_{x}^{\dagger} \pm i \hat{a}_{y}^{\dagger})/\sqrt{2}$ for left- and right-circular modes, act on the vacuum state $\vert 0 \rangle$.
These operators define the Heisenberg-Weyl algebra $\left[ \hat{a}_{j}, \hat{a}_{k}^{\dagger} \right] = \delta_{j,k}$ with $j,k=\pm$.
Assuming a harmonic oscillator in dimensionless polar configuration space, these modes correspond to the standard Laguerre-Gauss modes,
\begin{align}
    \begin{aligned}
        \left\langle r , \phi \vert  p, \ell \right\rangle =&~  \frac{1}{\sqrt{\pi }} (-1)^{p} \sqrt{\frac{ p!}{(p+\vert \ell \vert)!}} \, r^{\vert \ell \vert} e^{ - \frac{1}{2} r^{2}} \mathrm{L}_{p}^{\vert \ell \vert} (r^{2}) e^{i \ell \phi},
    \end{aligned}
\end{align}
written in terms of associated Laguerre polynomials, $\mathrm{L}_{p}^{\vert \ell \vert}(r^2)$, of the isotropic 2D harmonic oscillator \cite{Dennis2017}, 
\begin{align}
    \hat{H} = \frac{1}{2}  \hbar \omega \sum_{j=x,y} \left( \hat{p}_{j}^{2} +  \hat{q}_{j}^{2} \right),
\end{align}
where we use re-normalized dimensionless canonical variables, $ \hat{q}_{j} = \sqrt{m \omega / \hbar} \, \hat{x}_{j}$ and $\hat{p}_{j} = \sqrt{1 / (\hbar m \omega)} \, \hat{p}_{\hat{x}_{j}} = - i \partial_{q_{j}}  $, such that the annihilation (creation) operators are $\hat{a}_{j} = \left( \hat{x}_{j} + \partial/\partial x_{j} \right) / \sqrt{2}$ ($\hat{a}^\dagger_j=\left( \hat{x}_{j}-{{\partial}}/{{\partial x_{j}}} \right)/\sqrt{2}$) with $\hat{x}_{j} = \hat{x}, \hat{y} $.  

\begin{figure}
	\centering
	\includegraphics[scale=1]{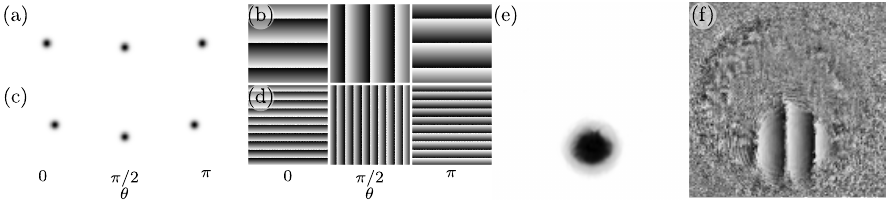}
	\caption{(a) and (c): Probability density function, $\vert \langle r,\phi \vert n_{-},\alpha \rangle \vert^{2}$.  (b) and (d): Phase, $\mathrm{arg}( \langle r,\phi \vert n_{-},\alpha \rangle)$, for Glauber-Sudarshan coherent states for the Heisenberg-Weyl algebra, $\vert n_{\pm} =0, \alpha \rangle$, with coherent parameter, $\alpha = \zeta e^{i \theta}$, amplitudes (a)-(b) $\zeta = 1$  and (c)-(d) $ \zeta = 3$ and different phase values $\theta = 0, \pi/2, \pi.$
    (e)-(f) Experimental classical intensity and phase for $\zeta = 3$ and $\theta = \pi/2$. } \label{fig:Fig1}
\end{figure}

Parceling the Hilbert space into subspaces with constant circular excitation number, $ n_{\pm} = \langle \hat{a}^{\dagger}_{\pm} \hat{a}_{\pm} \rangle  $, reveals an underlying Heisenberg-Weyl algebra.
The Glauber-Sudarshan coherent states \cite{Glauber1963,Sudarshan1963} for the subspace with null excitation,
\begin{align}
    \begin{aligned}
        \langle r, \phi \vert n_{\pm}= 0, \alpha \rangle =&~  \langle r, \phi \vert e^{ - \left( \alpha \hat{a}_{\pm}^{\dagger} - \alpha^{\ast} \hat{a}_{\pm}  \right)} \vert 0 \rangle \\
        =&~ \frac{1}{\sqrt{\pi}} e^{-\frac{\vert \alpha \vert^{2}}{2}} e^{- r \alpha e^{\pm i \phi}} e^{- \frac{1}{2}r^{2}},
    \end{aligned}
\end{align}
resembles a Gaussian function displaced by the radial displacement,
\begin{align}
    \begin{aligned}
    r_{0}(\alpha) =&~ \frac{\sqrt{\pi}}{2} e^{-\frac{\vert \alpha \vert^{2}}{2}} \left[ \left( \vert \alpha \vert^{2} +1 \right) \mathrm{I}_{0}\left( \frac{\vert \alpha \vert^2 }{2}\right) +
    \vert \alpha \vert^{2} \mathrm{I}_{1}\left( \frac{\vert \alpha \vert^2 }{2}  \right)\right], 
    \end{aligned}
\end{align}
in terms of the modified Bessel functions of the first kind, $\mathrm{I}_{j}(\vert \alpha\vert^{2}/2)$, controlled by the complex coherent parameter amplitude, $\alpha$. 
The coherent parameter phase controls the azimuthal position for the probability distribution center, Figs. \ref{fig:Fig1}(a) and \ref{fig:Fig1}(c).
The probability distribution is identical for the left- or right-circular operators, with their phases inverted.
As with any coherent state, its evolution keeps the coherent parameter amplitude but changes the phase, Figs. \ref{fig:Fig1}(b) and \ref{fig:Fig1}(d), resulting in the probability density rotating along a circular trajectory with radius $r_{0}(\alpha)$ around the dimensionless configuration space origin.
Figures \ref{fig:Fig1}(e) and  \ref{fig:Fig1}(f) show the experimental intensity and the phase.

\begin{figure}
	\centering
	\includegraphics[scale=1]{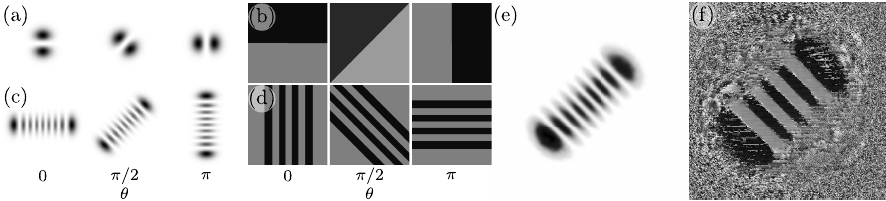}
	\caption{Same as Fig. \ref{fig:Fig1} for Gilmore-Perelomov coherent states for constant total excitation number subspaces, $\vert j; \zeta, \theta \rangle$, with coherent parameter amplitude $\zeta = \pi/2$ for Bargmann parameter (a)-(b) $j=1/2$ and (c)-(d) $j =4$ and different phase values $\theta = 0, \pi/2, \pi$.
    (e)-(f) Experimental intensity and phase for $ \zeta = \pi/2$, $\theta = \pi/2$, and $j = 4$.
    } \label{fig:Fig2}
\end{figure}

Parceling the Hilbert space into constant total excitation number, $ j = \langle \hat{a}^{\dagger}_{+} \hat{a}_{+} + \hat{a}^{\dagger}_{-} \hat{a}_{-} \rangle /2$, subspaces reveals an underlying symmetry described by the $su(2)$ algebra.
Its Gilmore-Perelomov coherent states \cite{Gilmore1972,Perelomov1972}, 
\begin{align}
    \begin{aligned}
        \left\langle r, \phi \vert j; \zeta , \theta \right\rangle =&~ \langle r, \phi \vert e^{ - \frac{\zeta}{2} \left( e^{-i \theta} \hat{a}^{\dagger}_{+} \hat{a}_{-} - e^{i \theta} \hat{a}_{+} \hat{a}^{\dagger}_{-} \right)} \vert j; 0 \rangle\\
        =&~ \sqrt{ \dfrac{(2j)!}{\pi} }  \left( \dfrac{- e^{  - i  \theta  } \tan \dfrac{\zeta}{2}}{1 + \left\vert \tan \dfrac{\zeta}{2} \right\vert^{2} } \right)^{j}  e^{-\frac{1}{2} r^{2} }
        \times \\ &~ \times
        \sum_{m=-j}^{j} \sqrt{\dfrac{ (j - \vert m \vert)! }{(j - m)! (j + m)! (j + \vert m \vert)! }} \times \\
        &~ \times \left(  - e^{- i \theta}  \tan \dfrac{\zeta}{2}\right)^{m} r^{2 \vert m \vert} \mathrm{L}_{j - \vert m \vert}^{2 \vert m \vert}(r^{2}) e^{i 2 m \phi}, 
    \end{aligned} \label{eq:SU2GP}
\end{align}
are given in terms of the complex coherent parameter, $\zeta e^{i \theta}$ with $\zeta \in [0, \pi ]$ and $\theta \in [0, 2 \pi )$.
The dimensionless configuration operators expectation values are null, $\hat{q}_{x} = \hat{q}_{y} = 0$.
Thus, the probability density function centers at the origin, exhibiting nodes  equal to twice the Bargmann parameter.
It maintains its shape, albeit a rotation governed by the coherent parameter phase, Figs. \ref{fig:Fig2}(a) and \ref{fig:Fig2}(c).
Figures \ref{fig:Fig2}(e) and \ref{fig:Fig2}(f) show the experimental intensity and the phase.

\begin{figure}
	\centering
	\includegraphics[scale=1]{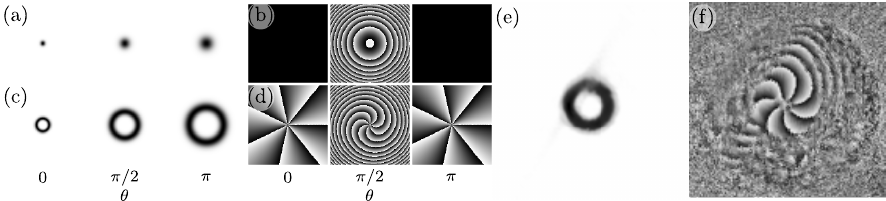}
	\caption{Same as Fig. \ref{fig:Fig1} for Gilmore-Perelomov coherent states for constant azimuthal number subspaces, $\vert k; \zeta, \theta \rangle_{\pm}$, with coherent parameter amplitude $\zeta = 1$ for Bargmann parameter (a)-(b) $k=1/2$ and (c)-(d) $k =4$ and different coherent phase values $\theta = 0, \pi/2, \pi$.
    (e)-(f) Experimental intensity and phase for $ \zeta = 1$, $\theta = \pi/2$, and $k=4$.
    } \label{fig:Fig3}
\end{figure}

Parceling the Hilbert space into constant azimuthal number, $k = ( \vert \langle \hat{\ell} \rangle \vert  +1)/2$,  subspaces reveals an underlying symmetry described by the $su(1,1)$ algebra. 
To distinguish positive and negative azimuthal number values sharing the same Bargmann index $k$, we introduce a new index.
The Gilmore-Perelomov coherent states,
\begin{align}
    \begin{aligned}
        \left\langle r, \phi \vert k; \zeta , \theta \right\rangle_{\pm} =&~   \langle r, \phi \vert e^{ - \frac{\zeta}{2} \left( e^{-i \theta} \hat{a}^{\dagger}_{+} \hat{a}^{\dagger}_{-} - e^{i \theta} \hat{a}_{+} \hat{a}_{-} \right)} \vert k; 0 \rangle_{\pm} \\
        =&~ \dfrac{1}{\sqrt{\pi (2k-1)!}} \dfrac{1 }{\sigma^{ 2k}} r ^{2k-1} e^{-\frac{r^2}{2 \sigma^{2}} } e^{ i \varphi_{\pm}(\phi; \theta) } 
        ,
   \end{aligned}
\end{align}
are given in terms of the complex coherent parameter, $\zeta e^{i \theta}$ with $\zeta \ge 0$ and $\theta \in [0, 2 \pi)$, the Gamma function, $\Gamma(2k)$, the Gaussian envelope waist,
\begin{align}
\sigma^{2} = \cosh \zeta - \cos \theta \sinh \zeta,
\end{align}
and the phase function, 
\begin{align} \varphi_{\pm}(\phi; \theta) =  \pm (2k -1) \phi -  \dfrac{\sin \theta \sinh \zeta}{\cosh \zeta - \cos \theta \sinh \zeta }   r^2,
\end{align}
up to a constant phase.
The Gaussian envelope waist decreases with an increasing real coherent parameter, confining the wave function near the origin. 
For complex parameters, it follows the coherent parameter phase increasing and decreasing in the ranges $\theta \in [0, \pi]$ and $ \theta \in [ \pi , 2 \pi)$, respectively.
The probability density function centers at the origin.
For the lowest Bargmann parameter, $k=1/2$, it is a Gaussian, Fig. \ref{fig:Fig3}(a), and a single ring, Fig. \ref{fig:Fig3}(c), for higher order Bargmann parameters, $k = 1, 3/2, 2, \ldots$.
The wave function phase for the lowest Bargmann parameter, $k=1/2$, takes a constant (annular) shape for real (complex) coherent parameters, Fig. \ref{fig:Fig3}(b), whereas for higher Bargmann parameters, $k = 1, 3/2, 2, \ldots$, it is a (swirl) vortex with topological order $2k-1$ for (complex) real coherent parameters, Fig. \ref{fig:Fig3}(d).
It shows a quadratic radial dependence and evolves between constant and annular patterns for $k=1/2$ or between the vortex and the swirl-vortex for all other Bargmann parameters,  Figs. \ref{fig:Fig3}(b)  Fig. \ref{fig:Fig3}(d).
Figures \ref{fig:Fig3}(e) and \ref{fig:Fig3}(f) show the experimental intensity and the phase.

\begin{figure}
	\centering
	\includegraphics[scale=1]{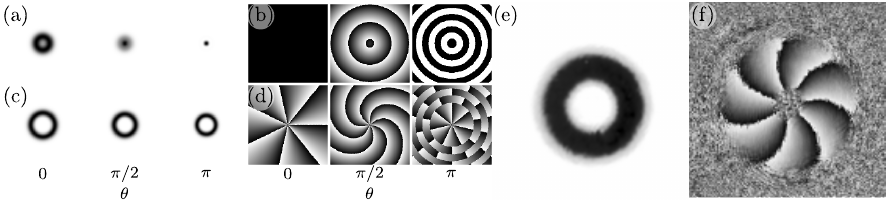}
	\caption{Same as Fig. \ref{fig:Fig1} for Barut-Girardello coherent states for constant azimuthal number subspaces, $\vert k; \xi \rangle_{\pm}$, with coherent parameter, $\xi = \zeta e^{i \theta}$, amplitude $\zeta = 1$ for Bargmann parameter (a)-(b) $k=1/2$ and (c)-(d) $k =4$ and different coherent phase values  $\theta = 0, \pi/2, \pi$.
    (e)-(f) Experimental intensity and phase for $ \zeta = 1$, $\theta = \pi/2$, and $k=4$.
    } \label{fig:Fig4}
\end{figure}

The Barut-Girardello coherent states \cite{Barut1971}, 
\begin{align}
    \begin{aligned}
       \left\langle r, \phi \vert k; \xi \right\rangle_{\pm} =&~ \frac{(-1)^{-k + 1/2} e^{-\xi}}{\sqrt{ \pi \, \mathrm{I}_{2k-1}(2 \vert \xi \vert)}} 
       \times \\ &~ \times 
       e^{-\frac{1}{2}r^{2}} \mathrm{J}_{2k - 1}\left( 2 \sqrt{- \xi} \, r \right) e^{ \pm i (2k -1) \phi},     \label{eq:SU11BG}
    \end{aligned}
\end{align}
takes the form of a Bessel-Gauss mode with complex argument scaled by the square root of the coherent parameter $\xi = \zeta e^{i\theta}$.
Its probability density distribution takes a Bessel shape for $k=1/2$ and a ring shape for higher Bargmann parameters. 
The minimum radius occurs for negative real values of the coherent parameter, Figs. \ref{fig:Fig4}(a) and \ref{fig:Fig4}(c).
The wave function phase shows a vortex with topological order $2k -1$ for complex coherent parameters whose phase is an integer multiple of $\pi$, and a swirl vortex for all other cases,  Figs. \ref{fig:Fig4}(b) and \ref{fig:Fig4}(d).
Figures \ref{fig:Fig4}(e) and \ref{fig:Fig4}(f) show the experimental intensity and the phase.

\begin{figure}
	\centering
	\includegraphics[scale=1]{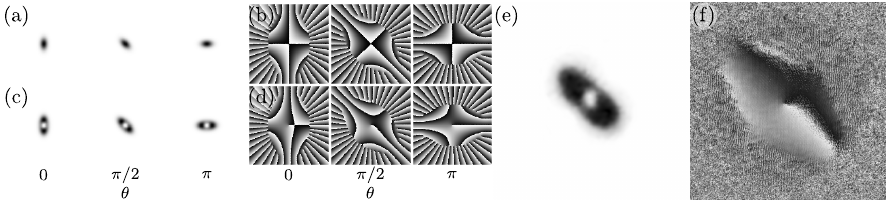}
	\caption{Same as Fig. \ref{fig:Fig1} for Gilmore-Perelomov coherent states for constant circular excitation number subspaces, $\vert k; \zeta, \theta \rangle_{n_{\pm}}$, with coherent parameter amplitude $\zeta = 1$ for Bargmann parameter (a)-(b) $k=1/4$ and (c)-(d) $k =3/4$ and different coherent phase values $\theta = 0, \pi/2, \pi$.
    (e)-(f) Experimental intensity and phase for $ \zeta = 1$, $\theta = \pi/2$, and $k=3/4$.} \label{fig:Fig5}
\end{figure}

Parceling the Hilbert space into constant circular excitation number, $n_{\pm}$, subspaces reveals additional representations, $k=1/4, 3/4$, of the  $su(1,1)$ algebra for each subspace.
Its Gilmore-Perelomov coherent states, 
\begin{align}
    \begin{aligned}
        \left\langle r, \phi \vert k; \zeta , \theta \right\rangle_{n_{\pm}} =&~ \dfrac{\sqrt{2k + \frac{1}{2}}}{\pi^{3/4}} \left( 1 -  \tanh^{2} \frac{\zeta}{2}  \right)^{k}  e^{-\frac{1}{2}r^{2}}
        \times \\ &~ \times
        \sum_{m=0}^{\infty} \sqrt{ \dfrac{\Gamma(2k + m) p_{\pm}!}{m! (p_{\pm} + \vert \ell_{\pm} \vert)!}} \times \\
        &~ \times \left( -e^{-i \theta} \tanh \frac{\zeta}{2} \right)^{m}   r^{\vert \ell_{\pm} \vert} \mathrm{L}_{p_{\pm}}^{\vert \ell_{\pm} \vert}\left( r^2 \right) e^{i \ell_{\pm} \phi},
   \end{aligned}
\end{align}
with radial and azimuthal numbers, 
\begin{align}
    \begin{aligned}
        p_{\pm} =&~ \dfrac{1}{2} \left[ n_{\pm} + 2 \left( k + m -\frac{1}{4} \right) - \vert \ell_{\pm} \vert \right] , \\
        \ell_{\pm} =&~ \mp n_{\pm} \pm 2 \left( k + m -\frac{1}{4} \right),
    \end{aligned}
\end{align}
are given in terms of the constant circular excitation number $n_{\pm}$.
For a given value of the Bargmann parameter, $k$, circular excitation numbers, $n_{+} = n_{-}$ such that $p_{+} = p_{-}$ and $\ell_{+} = \ell_{-}$, yield coherent states that are identical except for the phase term $e^{i  \ell_{\pm} \phi}$.
Thus, exploring one subspace is sufficient, the coherent state in the other subspace has its phase inverted.
The probability density function centers at the origin.
Numerical analysis suggests compression along the $q_{x}$ direction for real coherent parameters. 
For subspaces with Bargmann parameter $k=1/4$, the probability distribution shows $n_{\pm}$ nodes in the $q_{x}$ direction and no nodes in the $q_{y}$ direction.
Conversely, for $k=3/4$, it shows $n_{\pm}-1$ nodes in the $q_{x}$ direction at $q_{y} = 0$ and $n_{\pm}$ nodes at $q_{y} \neq 0$, with zero or one node in the $q_{y}$ direction depending on $q_{x}$ value.
The wave function phase shows a vortex with a hyperbolic-like pattern,  Figs. \ref{fig:Fig5}(b) and  \ref{fig:Fig5}(d), keeping their shape invariant.
Figures \ref{fig:Fig5}(e) and \ref{fig:Fig5}(f) show the experimental intensity and the phase.

\begin{figure}
	\centering	\includegraphics[scale=1]{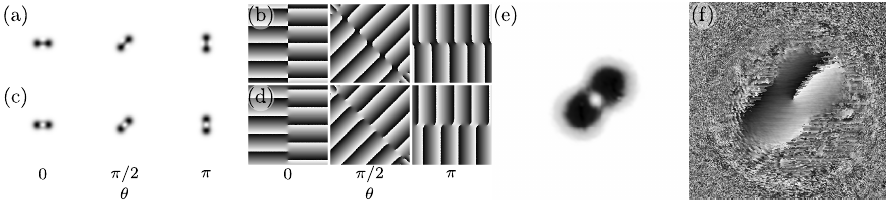}
	\caption{Same as Fig. \ref{fig:Fig1} for Barut-Girardello coherent states for constant circular excitation number subspaces, $\vert k; \xi \rangle_{\pm}$, with coherent parameter, $\xi = \zeta e^{i \theta}$, amplitude $\zeta = 1$ for Bargmann parameter (a)-(b) $k=1/4$ and (c)-(d) $k =3/4$ and different coherent phase values $\theta = 0, \pi/2, \pi$. (e)-(f) Experimental intensity and phase for $ \zeta = 1$, $\theta = \pi/2$, and $k=3/4$.
    } \label{fig:Fig6}
\end{figure}

The Barut-Girardello coherent states, 
\begin{align}
    \begin{aligned}
        \left\langle r, \phi \vert k; \xi \right\rangle_{n_{\pm}} = &~ \dfrac{\xi^{k - \frac{1}{2}} e^{-\frac{1}{2} r^{2}} }{\sqrt{\pi \, \mathrm{I}_{2k-1}(2 \vert \xi \vert)}} \sum_{m=0}^{\infty} \sqrt{ \dfrac{1 }{ m! \Gamma(2k + m )} } \xi^{m} 
        \times \\ &~ \times
        \sqrt{ \dfrac{ p_{\pm}! }{(p_{\pm}+\vert \ell_{\pm} \vert)!} } (-1)^{p_{\pm}} r^{\vert \ell_{\pm} \vert} \mathrm{L}_{p_{\pm}}^{\vert \ell_{\pm} \vert}\left( r^{2} \right) e^{i \ell_{\pm} \phi} ,
    \end{aligned}
\end{align}
shows that we need to explore just one of the subspaces defined by the circular excitation number, knowing that the coherent state in the other subspace has its phase inverted.
Their probability density function and wave function phase displays complex shapes, Fig.\ref{fig:Fig6}. 
Under evolution, both rotate with the coherent parameter phase,  Figs. \ref{fig:Fig6}(b) and \ref{fig:Fig6}(d), centered at the origin.  
Figures \ref{fig:Fig6}(e) and \ref{fig:Fig6}(f) show the experimental intensity and the phase.

\section{Methods}\label{sec11}

\begin{figure}
    \centering
    \includegraphics[width=0.95\linewidth]{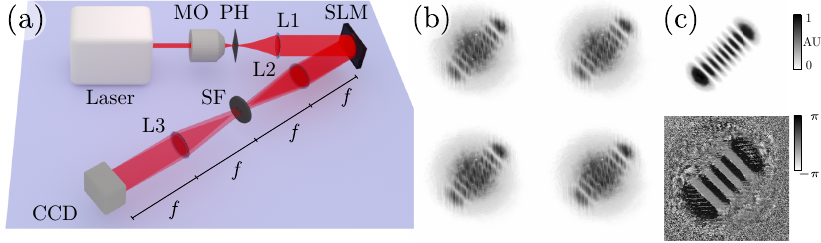}
    \caption{(a) Experimental setup.  Laser: He--Ne source; MO: Microscope Objective; PH: Pinhole; L1--L3: lenses; SLM: Spatial Light Modulator; SF: Spatial Filter; CCD: sensor. (b) Exemplary interference patterns.  (c) Recovered intensity (top) and phase (bottom).
    } \label{fig:Fig7}
\end{figure}

Figure \ref{fig:Fig7}(a) displays our experimental setup where a low--power HeNe laser illuminates the Spatial Light Modulator (SLM Holoeye Pluto VIS) after expansion and collimation.  
Utilizing a digital hologram \cite{Forbes2016} of our classical coherent modes, lenses L2 and L3 and an iris diaphragm (SF) filter and reconstruct the desired optical field in the first diffraction order.  
The CCD sensor (Nikon D3100) captures the optical field intensity in the back focal plane of L3.

We employed the four-frame technique \cite{Gaasvik2003}, measuring the interference pattern between the unknown beam and a reference wave, to determine the transverse phase of the beam.
The intensity $I_k(x,y) = a(x,y) + b(x,y)\cos[\varphi(x,y) + \alpha_k],$ where four phase shifts, $\alpha_k = (k-1)\pi/2$ with $k=1,2,3,4$, help compute the phase,
\begin{align}
	\varphi(x,y) = \text{atan}\left[\frac{I_4(x,y) - I_2(x,y)}{I_1(x,y) - I_3(x,y)}\right].
\end{align}
Our experimental setup has a dual purpose, generating and observing the field intensity and the phase via multiplexed holography \cite{Rosales-Guzman2017}; i.e., generating two beams, the desired optical mode along wave vector $\vec{k}_1$ and a reference beam in the $\vec{k}_2$ direction.  
Careful spatial filtering (SF) allows both beams to interfere in the CCD plane.  
The digital hologram includes necessary phase shifts, eliminating the need for any moving optical elements. 
Figure \ref{fig:Fig7}(b) provides some interference patterns, while Fig.\ \ref{fig:Fig7}(c) shows the recovered intensity (top) and the phase (bottom).

\section{Conclusion} 

In summary, we demonstrated how the Heisenberg-Weyl algebra provides coherent states in the configuration space, characterized by the Gaussian probability density functions that exhibit radial displacement and rotation during evolution.  
Further, the $su(2)$ algebra provides Gilmore-Perelomov (Bloch) coherent states that rotate around the origin while maintaining their shape. 
Interestingly, the $su(1,1)$ algebra, with two partitions, provides the Gilmore-Perelomov and Barut-Girardello coherent states. 
The first partition yields the Gaussian envelope probability density functions, transitioning from the Gaussian to single-ring shapes with increasing Bargmann parameter values. 
The second partition results in symmetrical probability density functions with unique vortex configurations.

Our work establishes a direct connection between the spatial parameters of classical optical beams and the dynamics of multiparticle systems, which is defined by the excitation mode of the electromagnetic field \cite{Sudarshan1963, Glauber1963b}.
These findings, together with recent attempts to develop orbital-angular-momentum beam splitters, might enable the use of spatial light modes to emulate multiparticle interference effects \cite{Hong1987, Zou2005}. 
This possibility has implications for optical and quantum information processing \cite{goodman2005introduction, Flamini2019}. 
In addition, our work contributes to ongoing efforts that aim to develop classical analogs of quantum systems. 
So far, extensive research has been directed toward the use of classical non-separability to explore complex entanglement dynamics \cite{Hashemi2015, Ndagano2017, Paneru2020, Qian2015}. 
However, our work strikes out in a new direction. 
It suggests that the spatial degree of freedom of an optical beam could be used to explore effects hosted by many-body systems.

\section*{Funding}
    M.P.M.R.: Fundaci\'on Carolina.
    O.S.M.L.: U.S. Department of Energy (DOE), Office of Science (SC), Program of Nuclear Physics, NP-QIS grant: DE-SC0023694.
    L.M.N.: Spanish MCIN, European Union Next Generation EU (PRTRC17.I1); Consejeria de Educacion, JCyL, QCAYLE, and MCIN PID2020-113406GB-I00 and RED2022-134301-T. 




\section*{References}

%

\end{document}